\documentclass[aps,showkeys,floatfix,twocolumn]{revtex4}

\usepackage[english]{babel}
\usepackage{amsmath}
\usepackage{graphicx}
\usepackage{microtype}

\newcommand {\sqrbrc}[1]{\left[#1\right]}
\newcommand {\brc}[1]{\left(#1\right)}
\newcommand {\avr}[1]{\left<#1\right>}
\newcommand {\abs}[1]{\left|#1\right|}
\newcommand {\txt}[1]{\text{#1}}
\newcommand {\ten}[1]{\mathbf{#1}}

\renewcommand {\vec}[1]{\mathbf{#1}}

\begin{document}

\title{Rotational relaxation time as unifying time scale for polymer and fiber drag reduction}
\date{\today}
\author{A.M.P. Boelens}
\author{M. Muthukumar}
\email{muthu@polysci.umass.edu}
\affiliation{Department of Polymer Science and Engineering, University of
Massachusetts Amherst, USA}

\begin{abstract}
Using hybrid Direct Numerical Simulation with Langevin dynamics, a comparison is
performed between polymer and fiber stress tensors in turbulent flow. The stress
tensors are found to be similar, suggesting a common drag reducing mechanism in
the onset regime for both flexible polymers and rigid fibers. Since fibers do
not have an elastic backbone, this must be a viscous effect. Analysis of the
viscosity tensor reveals that all terms are negligible, except the off-diagonal
shear viscosity associated with rotation. Based on this analysis, we identify
the rotational orientation time as the unifying time scale setting a new time
criterion for drag reduction by both flexible polymers and rigid fibers.
\end{abstract}

\keywords{}

\maketitle


\section{Introduction}
When a Newtonian fluid transitions from laminar flow to turbulent flow, changes
in pressure and velocity fields become chaotic, and eddies, coherent patterns of
flow velocity and pressure, start to form. As eddies break up into smaller
eddies, an energy cascade forms, which transports the kinetic energy of the flow to
smaller, and smaller time and length scales. Eventually, at the smallest scales,
also called the Kolmogorov scales, this kinetic energy gets dissipated into
heat, due to viscosity. Once a flow is turbulent, the energy cascade is
sustained through the turbulence regeneration cycle \cite{jimenez1999}. Drag
reduction is the phenomenon where, by either modifying the boundary conditions
on the wall \citep{choi1993,choi1994} or by adding additives
\citep{toms1948,pirih1972,lee1974,radin1975} to the flow, the turbulence
regeneration cycle is disrupted and the dissipation of turbulent kinetic energy is
reduced. Because of their effectiveness as drag reducing agents, polymers are a
popular type of additive \citep{toms1948}. By adding only a couple of parts per
million of certain polymers to a fluid, well below their overlap concentration
where polymer-polymer interactions are negligible, a drag reduction of up to 80 percent can be observed \citep{gampert1982}. Fibers
are another additive that also generate drag reduction \citep{radin1975}, and
one of the open questions in drag reduction is whether fibers and polymers share
the same drag reduction mechanism. Considering that fibers are simply very stiff
polymers, one could regard fiber drag reduction as a limiting case of polymer drag
reduction and make the assumption that they share the same drag reducing
mechanism. On the other hand, it has been reported that polymers have an onset criterion \citep{virk1967},
while fibers do not \citep{paschkewitz2004}. Additionally, it has been found
that fibers are not as effective drag reducing agents as polymers
\citep{paschkewitz2004}, and there is the viscosity \citep{lumley1969} versus
elasticity \citep{degennes1986} debate.

Because polymers are typically a lot smaller than the Kolmogorov length scale,
while their relaxation times overlap with the Kolmogorov time scale, the onset
of drag reduction for polymers is determined by a time criterion
\citep{lumley1969}. Fibers are not elastic and thus do not have an onset
criterion like flexible polymers do. However, their molecular weight does have an effect
on their drag reducing effect \cite{sasaki1992}, and a critical aspect ratio for
fibers has been found \cite{radin1975}. Since
fibers are also typically smaller than the Kolmogorov scale, rather than a length
scale, it can be expected that, like polymers, there is a fiber time scale
associated with their effectiveness as drag reducing agents.

The viscosity versus elasticity debate is centered around the question as to whether
polymer drag reduction is a local phenomenon caused by extensional viscosity, or
a non-local phenomenon caused by the transport of turbulent kinetic energy into
the polymer chain. In an extensional flow, when a critical shear rate is
reached, polymer coils stretch significantly compared to their equilibrium
state, which results in a significant increase of the elongational viscosity \citep{metzner1970}.
Replacing the inverse critical shear rate with the Kolmogorov time scale, this viscosity increase was proposed by
\citet{lumley1969} as a mechanism for drag reduction. The first to suggest that
elasticity is essential for drag reduction was \citet{degennes1986}. Based on
work by \citet{daoudi1978} he concluded that the elongational viscosity theory
could not be correct due to the absence of the coil-stretch transition for
polymers undergoing randomly fluctuating stresses in a turbulent velocity field,
and reasoned that drag reduction had to be the result of the elastic properties
of polymers instead \citep{tabor1986,sreenivasan2000}. Based on experimental
work, theory, and simulations, there is support for both theories \citep{moosaie2013}. Since fibers
do not have an elastic backbone their drag reducing effect is caused by
viscosity effects, and if elastic theory is right, it has to be concluded that
fibers and polymers have different drag reducing mechanisms. However, if the
drag reducing mechanism for polymers and fibers is the same, the conclusion has
to be that for polymers the drag reducing effect is also caused by viscosity.

The present work investigates the effect of elasticity on drag reduction by studying
the stress tensor, effective viscosity, and torque generated by different
polymers and fibers in turbulent pipe flow. To model the polymers and fibers, a
hybrid Direct Numerical Simulation/Langevin Dynamics approach is taken. This
way no closure models are needed to calculate the stress tensor for either the
polymers \citep{sureshkumar1997,zhou2003} or the fibers
\citep{paschkewitz2004,gillissen2007a}, and a direct comparison of the polymer
and fiber stress tensors is possible. While the polymers or fibers and solvent
are two way coupled, the number of molecules in the system is too small to
observe drag reduction in the velocity profile of the flow \citep{moosaie2013}.
This means that the results presented here are only applicable to the onset of drag
reduction and not the Maximum Drag Reduction (MDR) regime, the maximum amount of drag
reduction that can be observed in turbulent flow \citep{virk1975a}. 

The MDR regime has been previously addressed by \citet{lvov2004} and
\citet{benzi2005}. They have used the \citet{doi1986} stress tensor to model the
fibers and the \citet{giesekus1962} stress tensor, assuming that the flexible
polymers are in the Hookean regime \citep{bird1987}, for the elastic flexible polymers.  They
have reported that the physical origins of the stresses of flexible polymers and
rigid fibers are different, and have proposed different scaling laws for the
Reynolds stresses at the wall for elastic polymers and inelastic fibers.
Nevertheless, in the regime of the maximum drag reduction (MDR) asymptote, they
argued that both flexible polymers and rigid fibers give rise to an effective
viscosity and that this viscous effect is responsible for drag reduction in the
MDR regime for both flexible polymers and fibers.

In the present work, we focus on the onset regime of drag reduction, as pointed
out above. Both the flexible polymer and fiber are modeled as Finitely
Extensible Nonlinear Elastic (FENE) dumbbells with different Deborah numbers
distinguishing them. The main conclusion from the present Direct Numerical
Simulation/Langevin Dynamics approach is that both flexible polymer and  fiber
stress tensors have the same shape. This suggests that the drag reduction
mechanism for the flexible polymers and fibers are the same in the drag
reduction onset regime. Furthermore, analysis of the conformations of the
dumbbell model representing flexible chains shows that it is first stretched
into an anisotropic state with sufficiently large aspect ratio, thus
contributing to torque similar to fibers.  In agreement with \citet{sibilla2002}
and \citet{kim2008}, based on the work by \citet{lvov2004},
\citet{deangelis2004} and \citet{gillissen2007b}, we report that the
off-diagonal stress component is dominant in drag reduction. According to the
Kramers-Kirkwood equation \citep{kramers1944}, the off-diagonal stress component
is associated with rotation. Therefore, the dominant mechanism for drag
reduction in the onset regime is the rotation for both flexible polymers and
fiber. We also propose that the rotational relaxation time is the unifying time
scale between polymer and fiber drag reduction. 

\section{Model}
Different drag reduction methods act by reducing the momentum flux towards the
wall \citep{procaccia2008}. Since drag reduction is a wall phenomenon, time and
length scales are non-dimensionalized as:
\begin{align}
  t^{+} 
= 
  \frac{t \; u_{\tau}^{2}}{\nu_{1}}
& \;\;\txt{ and }\;\;
  x^{+}
= 
  \frac{x \; u_{\tau}}{\nu_{1}}.
\end{align}
In the above equations $u_{\tau} = \sqrt{{d_{1}}/\brc{4 \rho_{1}} \abs{{\Delta p}/{\Delta x}}}$
is the friction velocity, and $\nu_{1}$ the kinematic viscosity \citep{virk1975a}. 
The subscript $1$ is used to indicate that the variables
describe the solvent, while the polymer and fiber variables have a subscript $2$. $d_{1}$
is the diameter of the pipe, $\rho_{1}$ the density of the solvent, and $\Delta
p / \Delta x$ is the pressure gradient. All variables in this paper are in $+$
units, i.e. non-dimensionalized with the friction velocity and kinematic
viscosity, but for improved readability the $+$ superscript has been omitted. In
non-dimensional form, the Navier-Stokes equation describing the momentum balance
of the solvent is:
\begin{equation}
  \frac{\partial \vec{u}_{1}}{\partial t}
+ \vec{u}_{1} \cdot \nabla \vec{u}_{1}
=
- \nabla p
+ \nabla^2 \vec{u}_{1}
+ \vec{f}_{2}
\label{eqn:NavierStokesWallUnits}
\end{equation}
and conservation of mass is guaranteed by the continuity equation. In the above
equation, $\vec{u}$ is the velocity, $t$ is time, $p$ is the
pressure, and $\vec{f}_{2}$ is the polymer dumbbells acting on the solvent.
Polymers and the solvent are two-way coupled, i.e. both the solvent acting on the
polymers, and the resulting reactive force are accounted for. To minimize the
number of variables, gravity is neglected.

To be able to describe the forces of the polymer dumbbell back onto the solvent,
the polymer dumbbells are described by Langevin dynamics.  Because
the dominant time scale for polymer drag reduction is the longest relaxation
time \citep{lumley1969}, they are modeled as Finitely Extensible Nonlinear
Elastic (FENE) dumbbells \citep{bird1987}, and their longest relaxation time is the Zimm
relaxation time. The two beads of the dumbbell are called $A$ and $B$, and the
drag force on the beads is assumed to be Stokes drag. Polymer-polymer
interactions are neglected. Writing the equation of motion for bead $A$ in wall
units gives:
\begin{multline}
  \tau_{2} \ddot{\vec{x}}_{2,A}
= 
- \brc{\dot{\vec{x}}_{2,A} - {\vec{u}}_{1,A}} \\
- \frac{1}{\txt{De} \vphantom{\brc{\vec{x}_{2,AB}/\vec{x}_{2,\txt{Max}}}^{2}}} \;
  \frac{\vec{x}_{2,AB}}{1 - \brc{\vec{x}_{2,AB}/\vec{x}_{2,\txt{Max}}}^{2}}
+ f_{R} \brc{t}
\end{multline}
with $\vec{x}_{2,AB} = \brc{\vec{x}_{2,B} - \vec{x}_{2,A}} -
\vec{x}_{2,0}$. $\vec{x}_{2}$ is the position of a bead, $\vec{x}_{2,0}$ is the equilibrium
distance between beads $A$ and $B$, $\vec{x}_{2,\txt{Max}}$ is the maximum
extension, and $\vec{u}_{1,A}$ is the fluid velocity at the position of bead
$A$. Dots signify derivatives with respect to time. The random force, $f_{R}
\brc{t}$, is zero on average, and each hit by a solvent molecule is assumed
to be independent from all others. For the fibers, the spring force is left out
of their equation of motion and the beads are kept at fixed distance using the
RATTLE algorithm \citep{andersen1983}.

The simulations are modeled after a system of polyethylene glycol (PEG) in
water, and the following non-dimensional numbers result from making the above
equations dimensionless. The friction Reynolds number:
\begin{equation}
  \txt{Re}_{\tau}
= 
  \frac{d_{1} u_{\tau}}{\nu_{1}}
=
  560
\end{equation}
corresponds to a bulk Reynolds number of $\txt{Re} = 8800$, and is equivalent to
the non-dimensional diameter of the pipe. A constant friction Reynolds number
implies a constant pressure gradient, and variable bulk velocity. The Deborah
number: 
\begin{equation}
  \txt{De}
=
  \tau_{Z}
=
  0, 1, 10
\end{equation}
is defined as the ratio of the characteristic time scales of the polymers and
the solvent, and is a measure for polymer elasticity. Since the
characteristic time scale of the fluid in wall units is equal to one, the
Deborah number is equal to the Zimm relaxation time in wall units,
$\tau_{Z}$. $\txt{De}=1$ defines the onset of drag reduction, $\txt{De}=10$
is a value well within the drag reduction regime, and $\txt{De}=0$ is the value
for fibers. The particle relaxation time:
\begin{equation}
  \tau_{2}
=
  \frac{1}{\rho^{*} {d^{*}}^{2}} \frac{\txt{Re}_{\tau}^{2}}{18}
= 
  1.789 \cdot 10^{-3}
\end{equation}
with $\rho^{*} = \rho_{1} / \rho_{2}$, and $d^{*} = d_{1} / d_{2}$, is a
measure of how sensitive a bead is to velocity fluctuations in the
fluid. The last dimensionless group, the diffusion constant, equals: 
\begin{equation}
  D
=
  \frac{\brc{k_{B} T}^{+}}{\zeta}
=
  8.133 \cdot 10^{-4}
\end{equation}
with $\brc{k_{B} T}^{+} = k_{B} T {u_{\tau}}/{\rho_{1} \nu_{1}^{3}}$, $k_{B}$ 
the Boltzmann constant, $T$ the temperature, and $\zeta = 3 \pi
{\txt{Re}_{\tau}}/{d^{*}}$ the non-dimensional friction factor. This number
determines whether diffusion or advection is dominant. With changing Deborah
numbers, the molecular weight of the dumbbells has been kept constant, which
results in the following equilibrium lengths in order of increasing
Deborah number: $\vec{x}_{2,0} = 4.000 \cdot 10^{-1}$, $\vec{x}_{2,0} = 2.836 \cdot
10^{-2}$, and $\vec{x}_{2,0} = 6.110 \cdot 10^{-2}$. The maximum extensions
are: $\vec{x}_{2,\txt{Max}} = 1.000 \cdot \vec{x}_{2,0}$,
$\vec{x}_{2,\txt{Max}} = 14.10 \cdot \vec{x}_{2,0}$, and
$\vec{x}_{2,\txt{Max}} = 6.546 \cdot \vec{x}_{2,0}$. The number of
dumbbells in the system is $N_{2} = 9.600 \cdot 10^{ 5}$. Keeping the molecular
weight constant is equivalent to the experiments performed by \citet{virk1975b}
on drag reduction by rod-like and coiled polyelectrolytes. 

The code solves the Navier-Stokes equations in cylindrical coordinates using
Direct Numerical Simulation, and is based on work by \citet{eggels1994}.  $r$, $\phi$, and
$z$, are the radial, angular, and streamwise directions, respectively, and $u$,
$v$, and $w$, are the corresponding velocity components. The code is a 4th order
predictor-corrector finite volume code working on a non-homogeneous staggered
grid with Leap-Frog time stepping. Bead tracking uses the velocity Verlet
algorithm, and is based on work by
\citet{boelens2007}. Simulations are run on a grid of $128 \times 256 \times
256$. This is a coarse mesh, but the results are expected to hold for higher
resolutions \citep{choi1994}. More detailed information about the code can be
found in \citet{boelens2012}.

\section{Results}

\begin{figure}
\includegraphics[width=0.45\textwidth]{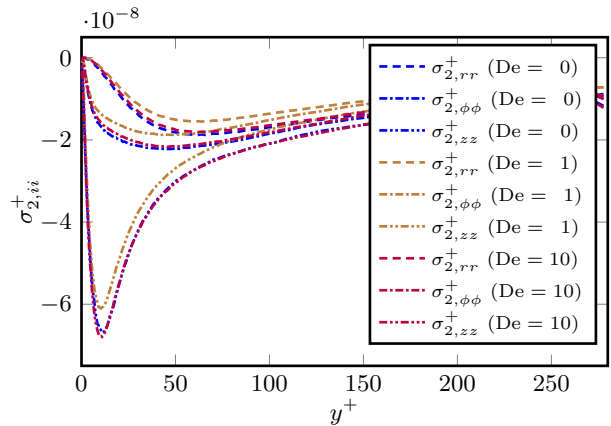}
\caption{\small Diagonal components of the polymer and fiber stress tensor. \label{fig:TauPii}}
\end{figure}
\begin{figure}
\includegraphics[width=0.45\textwidth]{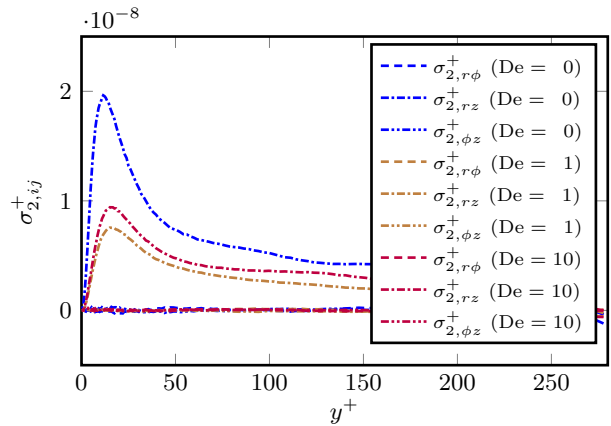}
\caption{\small Off-diagonal components of the polymer stress tensor. \label{fig:TauPij}}
\end{figure}

In Figures \ref{fig:TauPii} and \ref{fig:TauPij} the diagonal and off-diagonal
components of the polymer and fiber stress tensor, $\sigma_{2}$, are shown
as functions of the dimensionless distance from the wall. Taking the difference
in coordinate systems into consideration, the different components of the stress
tensor are in agreement with literature \cite{gillissen2007b}. It can be
observed that dumbbells with Deborah number $\txt{De} = 0$ have the largest,
$\txt{De} = 10$ the second largest, and $\txt{De} = 1$ the smallest stress
tensor components. This is consistent with the results of \citet{virk1975b}, who
found that for low concentrations in the onset regime rod-like polyelectrolytes
have a stronger drag reducing effect than coiled polyelectrolytes with the same
molecular weight. In addition, the behavior of the polymer stress tensors is
also as expected, because the dumbbell with $\txt{De} = 1$ was parameterized to be
the dumbbell at which drag reduction onset occurs, and thus polymer stresses are
the weakest. Comparing the stress tensor for $\txt{De} = 0$ with the other two
stress tensors for $\txt{De} = 1$ and $\txt{De} = 10$, it can be seen that, apart
from different values for the maxima and minima, the different stress tensor
components have exactly the same shape. Since the stress tensor describes the
full interaction of the dumbbells with the solvent, this suggests that the polymers
and fibers share the same drag reducing mechanism in the drag reduction onset
regime. In addition, since the elastic theory does not apply to fibers, it can be
concluded that drag reduction is a local phenomenon and is caused by viscous
effects. This is in agreement with \citet{gillissen2007b}. After recognizing that the r axis points to the wall while the y axis points out of the
wall, and considering whether the forces are on the beads or on the solvent, our results of the stress tensor are completely consistent with
Fig.3 of \citet{gillissen2007b}. On the other hand, our results are different from  \citet{lvov2004}
and \citet{benzi2005}. In their work, they use the \citet{doi1986} stress tensor
for the fibers and the \citet{giesekus1962} stress tensors for the
elastic polymers. In their analysis they find a coupling between the diagonal and off-diagonal
components of the conformation tensor which is linear for elastic polymers and
quadradic for rod-like polymers. \citet{benzi2005} propose a
different scaling for the Reynolds stresses at the wall for elastic and rod-like
polymers. Since the velocity profile at the wall in $+$ units is expected to be independent of
the details of the drag reduction agent \cite{procaccia2008}, this means that
the stress tensors for our rigid and elastic dumbbells should show different
scaling as well. We do not observe this difference in scaling in our
simulations. A possible origin of this difference is that the Giesekus tensor assumes that the
polymers are in the Hookean regime \citep{bird1987}, while in our system the
polymers are modeled as Finite Extensible Nonlinear Elastic dumbbells. Another
difference is that the work of \citet{lvov2004} and \citet{benzi2005} concerns
the Maximum Drag Reduction (MDR) regime, while our simulations are in the onset
regime.
Furthermore, our results show that the stress for rod-like polymers is higher
than that for flexible polymers. This is in agreement with \citet{virk1975b},
where polyelectrolyte chains in salt-free conditions (rod-like) and salty
conditions (coil-like) were investigated.

\begin{figure}
\includegraphics[width=0.45\textwidth]{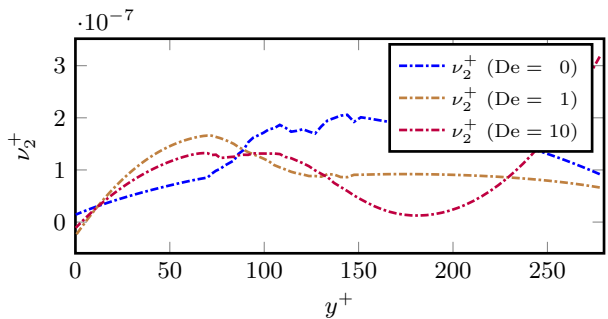}
\caption{Effective polymer viscosity. \label{fig:nuPij}}
\end{figure}

A follow up question that one can ask is where the effective viscosity
originates from. Both polymers and fibers are known to show a large increase in
viscosity in extensional flow, associated with the diagonal components of the
stress tensor \citep{metzner1970}, but there is also the shear viscosity which
is associated with the off-diagonal stress tensor component and rotation. To
analyze this question, we evaluate the contribution of the effective
viscosity to the momentum balance in its most general form:
\begin{equation}
  \nabla \cdot \sigma_{2}
=
  2 \nabla \cdot \brc{\ten{\nu}_{2} \mathbin{:} \ten{s}_{1}}
\end{equation}
with $\ten{s}_{1} = 1/2 ({\nabla \vec{u}_{1}}^{T} + \nabla \vec{u}_{1})$
the rate-of-strain tensor, and $\ten{\nu}_{2}$ the fourth order viscosity
tensor. Performing a Reynolds decomposition on this equation and taking into
account the symmetries of our system gives:
\begin{equation}
  \nabla \cdot \sigma_{2}
=
  \nabla \cdot \sqrbrc{
    \tilde{\nu}_{2} (r) \frac{\partial \avr{w_{1}}}{\partial r}
  + 2 \avr{\ten{\nu}_{2}'  \mathbin{:} \ten{s}_{1}'}    
  }.
\end{equation}
Here $'$ denotes the fluctuating part and $\avr{}$ the average with:
\begin{equation}
  \tilde{\nu}_{2} (r)
=
  \left[ 
  \begin{array}{ccc}
  \tilde{\nu}_{2,rr} (r) & 0                                        & \tilde{\nu}_{2,zr} (r) \\
  0                              & \tilde{\nu}_{2,\theta\theta} (r) & 0             \\
  \tilde{\nu}_{2,zr} (r) & 0                                        & \tilde{\nu}_{2,zz} (r)
  \end{array} 
  \right].
\end{equation}
Further expanding the above equation gives the following contributions to the Navier-Stokes equations:
\begin{widetext}
\begin{alignat}{3}
  \brc{\nabla \cdot \sigma_{2}}_{r} & = 
    \frac{1}{r} \frac{\partial}{\partial r} \sqrbrc{
    r \tilde{\nu}_{2,rr} (r) 
    \frac{\partial \avr{w_{1}}}{\partial r}
  }
- \frac{\tilde{\nu}_{2,\theta\theta} (r)}{r} \frac{\partial \avr{w_{1}}}{\partial r}
&& + 2 \brc{\nabla \cdot \avr{\ten{\nu}_{2}'  \mathbin{:} \ten{s}_{1}'}}_{r} \\
  \brc{\nabla \cdot \sigma_{2}}_{\theta} & =
  \phantom{a}
&& + 2 \brc{\nabla \cdot \avr{\ten{\nu}_{2}'  \mathbin{:} \ten{s}_{1}'}}_{\theta} \\
  \brc{\nabla \cdot \sigma_{2}}_{z} & =
    \frac{1}{r} \frac{\partial}{\partial r} \sqrbrc{
    r \tilde{\nu}_{2,zr} (r) 
    \frac{\partial \avr{w_{1}}}{\partial r}
  }
&& + 2 \brc{\nabla \cdot \avr{\ten{\nu}_{2}'  \mathbin{:} \ten{s}_{1}'}}_{z}.
\end{alignat}
\end{widetext}
These equations show how, in addition to
viscosity-velocity correlations, both the diagonal and off-diagonal components
of the stress tensor act on the solvent. While it has been shown that at least
at the Maximum Drag Reduction limit all components of the polymer stress tensor
are coupled \cite{lvov2005}, one can expect that by leaving out different terms
in the above equations their contribution to drag reduction can be investigated.
Based on work by \citet{lvov2004} the results of this can already be found in
literature \citep{deangelis2004,gillissen2007b}, where the above viscosity
tensor is replaced by a scalar viscosity function $\hat{\nu}_{2} (r)$.
The contribution to the Navier-Stokes equations can then be written as:
\begin{equation}
  \nabla \cdot \sigma_{2}
=
  \nabla \cdot \brc{\hat{\nu}_{2} (r) \ten{s}_{1}},
\end{equation}
which, after Reynolds decomposition, gives a contribution of the form:
\begin{align}
  \brc{\nabla \cdot \sigma_{2}}_{r} & = 0 \\
  \brc{\nabla \cdot \sigma_{2}}_{\theta} & = 0 \\
  \brc{\nabla \cdot \sigma_{2}}_{z} & = 
  \frac{1}{r} \frac{\partial}{\partial r} \sqrbrc{r
  \hat{\nu}_{2} (r) \frac{\partial \avr{w_{1}}}{\partial r}}.
\end{align}
The above equations only contain the off-diagonal shear viscosity component and
none of the extensional viscosity or viscosity-velocity fluctuations.
Calculating the fiber viscosity function and using this as an input in a new
simulation, all the characteristics of the drag reduced flow can be recovered
\citep{gillissen2007b}. \citet{deangelis2004} showed that a viscosity gradient
at the wall is also able to reproduce the characteristics of a polymer drag
reduced flow. This means that not only fluctuations can be ignored
\citep{gillissen2007b}, but also shows that the diagonal components of the
stress tensor, and thus extensional viscosity, can be neglected.

Figure \ref{fig:nuPij} shows the effective polymer and fiber viscosities
calculated from the stress tensor. Their shape of a gradient at the wall
and a plateau in the center is consistent with the viscosity profile used by
\citet{deangelis2004}.

\begin{figure}
\includegraphics[width=0.45\textwidth]{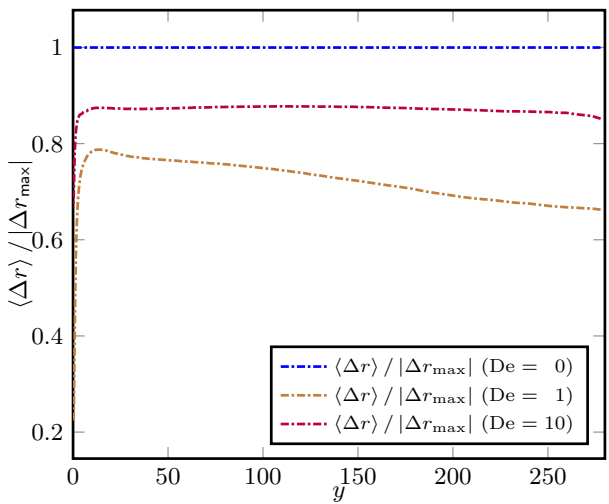}
\caption{Average relative length of the end-to-end vector of the polymer dumbbells.
\label{fig:endToEnd}}
\end{figure}
To further explore the idea of polymers and fibers creating drag reduction by
rotation, the average relative length of the end-to-end vector of all dumbbells
normalized by their maximum extensions are shown in Figure \ref{fig:endToEnd}.
Because the molecular weight is kept constant between the different simulations,
the maximum extension is $1$ for all cases. The dumbbell with a Deborah number
of $\txt{De} = 0$ is always fully extended. Because it represents the most
elastic dumbbell with the longest relaxation time, the dumbbell with a Deborah
number of $\txt{De} = 10$ gets stretched further than the dumbbell with a Deborah
number of $\txt{De} = 1$. This is in agreement with the results shown in Figure
\ref{fig:TauPij}, which showed that $\txt{De} = 0$ displays the largest amount of
drag reduction followed by Deborah numbers $\txt{De} = 10$ and $\txt{De} = 1$.
Both elastic dumbbells are stretched the most when close to the wall where gradients
are the largest, and relax towards the center of the pipe. That the amount of
drag reduction is proportional to the average relative length of the end-to-end
vector of the fibers and dumbbells, i.e. their moment arm, and not to the amount
of turbulent kinetic energy that can be stored in the backbone, is an
additional indication that polymer rotation is essential for drag reduction in
the onset regime.

\begin{figure}
\includegraphics[width=0.45\textwidth]{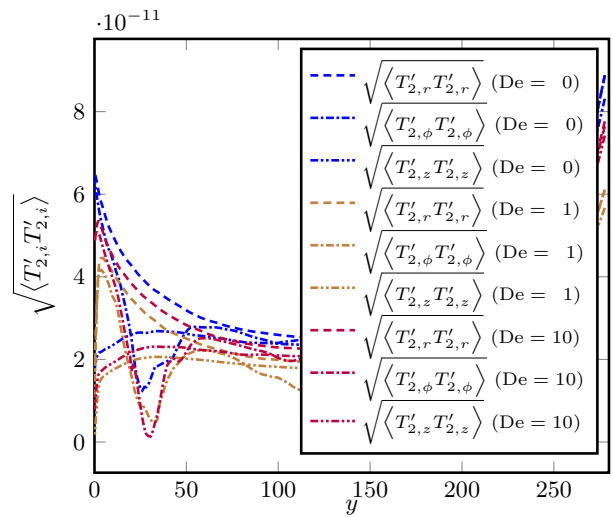}
\caption{Standard deviation of the torque exerted by the polymer dumbbells on the
solvent. \label{fig:Torque}}
\end{figure}
In addition to the end-to-end vector, one can also look into the standard
deviation of the torque, which is shown in Figure \ref{fig:Torque}. The torque
on the solvent is defined as: 
\begin{equation}
  T_{2}
=
- \frac{n}{2 V} \avr{\Delta \vec{x}_{2} \times \Delta {\vec{F}_{2}^{h}}}
\end{equation}
where $n$ is the number of polymer dumbbells in volume $V$, $\Delta
\vec{x}_{2} = \vec{x}_{2,A}-\vec{x}_{2,B}$ is the moment arm (i.e. the end-to-end
vector), $\Delta {\vec{F}_{2}^{h}} = {\vec{F}_{2,A}^{h}} -
{\vec{F}_{2,B}^{h}}$, with $\vec{F}_{2,i}^{h}$ the hydrodynamic drag force
on bead $i$, and $\avr{}$ indicates an ensemble average. Because of the
symmetry of the polymer stress tensor, all components of the torque
vector on average are zero. By looking at the standard deviation of the torque, it is
possible to gain more insight into polymer/fiber-solvent interactions. Away from the
wall the standard deviation of each torque component is about the same for each
polymer, indicating that they are freely tumbling around in the bulk. However,
as the wall is approached, the different components diverge. The torque has
three contributions: i) the length of the moment arm or end-to-end vector, ii) the magnitude of the
hydrodynamic forces, and iii) the wall blocking rotation. For the radial
component, the wall does not block rotation, and the arm is maximized due to
alignment in the streamwise direction. This results in a strong increase of the
radial torque fluctuations at the wall. In the streamwise direction, on the other
hand, the wall is blocking full rotation, which gives fluctuations which decline
monotonically to zero at the wall. For the angular torque fluctuations, the
component associated with the shear viscosity, close to the wall
rotation is blocked, so the fluctuations go to zero. However, further away from
the wall, the maximum moment arm causes a large increase in the torque fluctuations.
This shows that, although elasticity is not strictly necessary for drag
reduction, the coil-stretch transition is important for polymer drag reduction
because it generates a moment arm. Another way of looking at this
is that a polymer coil without any stretching can be thought of as a rough sphere which has no drag
reducing effect. Fibers, on the other hand, because of their aspect ratio, always
have a drag reducing effect. Polymer elasticity allows the roughly spherical coil
to stretch into an anisotropic state with an aspect ratio larger than unity. It is
not the elasticity that makes a polymer drag-reducing, but the fact that it can
transform into anisotropic conformations. This is consistent with the results of
\citet{sibilla2002} who suggested that shear viscosity
cannot be neglected, and \citet{kim2008}, who found that counter-torque by
polymers suppresses the formation of hairpin vortices at the wall.

\section{Conclusions}
To summarize, analyzing the polymer stress tensor it can be observed that
polymer and fiber stress tensors show the same characteristics. This indicates
that, at least in the low drag reduction regime, polymers and fibers show the
same drag reduction mechanism. Because fibers cannot store turbulent kinetic
energy in their backbone, this mechanism has to be caused by viscosity
effects. We find that the viscous effect arises from rotational motion of fibers and partially stretched 
flexible chains, which is taken as the unifying drag reduction mechanism in the onset regime. 
Although viscous effects have been previously attributed to drag reduction  in 
the MDR asymptote regime by \citet{lvov2004} and \citet{benzi2005}, they have argued that 
the physical origins of drag reduction by flexible polymers and rigid fibers are different. The difference in scaling
between elastic and rod-like polymers suggested by \citet{benzi2005} is not
observed in our simulations. It must be emphasized that the conclusions of
\citet{benzi2005} deal with the regime of maximum drag reduction (MDR)
asymptote, whereas our present work is in the onset regime of drag reduction. In
addition, \citet{benzi2005} assume the polymers are in the Hookean regime while
our dumbbells are modeled as Finite Extensible Nonlinear Elastic (FENE) springs.
Our result in the onset regime is qualitatively different from the arguments in 
\citep{procaccia2008} that for flexible polymers the main source of interaction with turbulent 
fluctuations is the stretching of polymers by the fluctuating shear and for rod-like polymers 
dissipation is only taken as the skin friction along the polymer. We find that the molecular rotation is 
the microscopic mechanism for the onset of drag reduction. 
By analyzing the different contributions to the effective viscosity tensor, based
on work by \citet{lvov2004}, \citet{deangelis2004}, and \citet{gillissen2007b},
it is found that all terms can be neglected except for the off-diagonal
component associated with polymer and fiber rotation. To further explore the
idea of polymer and fiber rotation being important in the onset of drag
reduction regime, polymer and fiber torque fluctuations are investigated. The
results suggest that the reason that the coil stretch transition is important
for polymer drag reduction is that it generates a moment arm. This is consistent
with the results of \citet{kim2008}.

Based on the findings in this work, we propose the rotational orientation time
\citep{kirkwood1951,muthukumar1983} as a time criterion for drag reduction. 

\begin{acknowledgments}
Many thanks to Prof. Perot at the Mechanical Engineering department of the
University of Massachusetts, Amherst for many stimulating discussions and
feedback. This research was supported by NSF Grant No. DMR-1404940 and AFOSR
Grant No. FA9550-14-1-0164.
\end{acknowledgments}


\end{document}